\documentclass[prl,twocolumn,epsf]{revtex4}

\input epsf

\newcommand{\be}[1]{\begin{equation}}
\newcommand{\ee}[1]{\end{equation}}
\begin{document}
\title{Eigenmodes and thermodynamics of a Coulomb chain in a harmonic
  potential} 
\author{Giovanna Morigi$^1$ and Shmuel Fishman$^2$} 
\address{$^1$ Department of Quantum Physics, University of Ulm,
D-89069 Ulm, Germany\\ 
$^2$ Department of Physics, Technion, 32000 Haifa, Israel} 

\date{\today}

\begin{abstract}
The density of ions trapped in a harmonic potential in
one dimension is not uniform. Consequently the eigenmodes are not
phonons. We calculate the long wavelength modes in the continuum
limit, and evaluate the density of states in the short wavelength
limit for chains of $N\gg 1$ ions. 
Remarkably, the results that are found analytically in the thermodynamic limit
provide a good
estimate of the spectrum of excitations of small chains down to few tens of
ions. 
The spectra are used to compute the thermodynamic functions of the chain.
Deviations from extensivity of the thermodynamic
quantities are found. An analytic
expression for the critical transverse frequency determining the
stability of a linear chain is derived.
\end{abstract}
\maketitle

Cold atomic gases are systems allowing an extremely high degree of control,
opening new directions and challenges both for experiments and
theory. The realization of low dimensional ultracold gases enables one to
design and investigate, among others, systems with peculiar excitations 
spectra and thermodynamic properties~\cite{BEC1d,Habs}. In this context, Coulomb
crystals constitute a unique form of an ordered structure, formed by cold ions
in a confining potential, that balances the Coulomb repulsion~\cite{DubinRMP}.
The ions vibrate about fixed positions in
analogy to the situation in an
ordinary solid, while the interparticle distance
is usually of the order of several micrometers, constituting an extremely
rarefied type of condensed matter~\cite{Drewsen03,NISTBoulder,MPQ,Raizen}. 
Variation of the potential
permits to control the crystal shape as well as 
the number of ions, allowing to explore crystals
of very different sizes, thus offering the opportunity of studying
the transition from few particles to mesoscopic systems
dynamics~\cite{DubinRMP}. Besides, these structures 
provide promising applications for spectroscopy~\cite{Werth}, 
frequency standards~\cite{Here2}, study and control of chemical
reactions~\cite{DrewsenChemistry}, and quantum information
processors~\cite{Cirac95,Steane,QLogic:2}.    

In this letter we investigate the dynamics of Coulomb chains, i.e.\
one-dimensional structures obtained by means of strong transverse
confinement and that usually consist of dozens of ions localized 
along the trap axis~\cite{Raizen,MPQ}. The equilibrium charge distribution for
such chains is {\it not} uniform~\cite{Dubin97,Schiffer03}. This is in
contrast to the three-dimensional ordering, where the density is uniform and
the eigenmodes are phonons. In this letter, the chain
excitations are calculated. These are fundamentally different from the phonons
in solids because of the non-uniformity of the ion distribution and the
long range of the Coulomb interaction. We derive
some thermodynamic functions of the chain, and define a specific
thermodynamic limit. Interestingly,
the linear ion chain differs from systems that are 
traditionally studied in the framework of statistical physics, since 
the thermodynamic quantities are not extensive. This behaviour
is due to the strong correlation and to the crystal
dimensionality. It manifests, for instance, in the dependence of the specific
heat on the number of ions, as we show below. The detailed analytical
derivation of the results presented here
will be published elsewhere~\cite{Morigi}.

We consider $N$ ions of mass $m$ and charge $Q$, which are confined by a
harmonic trap of cylindrical symmetry and are crystallized along
the trap axis at the 
positions $(x_j^{(0)},0,0)$, at which the harmonic force and
the Coulomb force, exerted by the other ions, balance. At sufficiently
low temperatures the vibrations around these points can be considered
harmonic and described by~\cite{James,Kielpinski} 
\begin{eqnarray} 
\label{Eq:ax}
&&\ddot{q}_i+\nu^2q_i=-I[x_i^{(0)},q_i]\\
&&\ddot{y}_i+\nu_t^2y_i=\frac{1}{2}I[x_i^{(0)},y_i]
\label{Eq:t}
\end{eqnarray}  
where $q_i=x_i-x_i^{(0)}$ and $y_i$ are the displacements
in the axial and transverse directions (the equations for $z_i$ are similar to
the ones for $y_i$), while the harmonic confinement is
characterized by the axial and transverse frequencies $\nu$, $\nu_t$, 
respectively ($\nu_t\gg
\nu$). The coupling matrix is
\begin{equation}
\label{I}
I[x_i^{(0)},w_i]=\frac{2Q^2}{m}\sum_{j\neq i}\frac{1}{|x_i^{(0)}-x_j^{(0)}|^3}(w_i-w_j)\\
\end{equation}
with $w_i=q_i,y_i,z_i$. Equations~(\ref{Eq:ax}-\ref{Eq:t})
describe a system of coupled oscillators, with long range interaction and 
position-dependent coupling strength. In what follows we calculate the
eigenmodes of this system. For this we assume $w_i(t)=\int {\rm e}^{{\rm
    i}\omega t}\tilde{w}_i(\omega){\rm d}\omega/2\pi $. To simplify notations we replace
$\tilde{w}_i(\omega)$ by $w_i$. This results in equations for the eigenmodes
of frequency $\omega$ that are similar to (\ref{Eq:ax}-\ref{Eq:t}), but
with $\ddot{w}_i$ replaced by $-\omega^2w_i$. The problem reduces to 
the eigenvalue equation
\begin{equation}
\label{I:2}
I[x_i^{(0)},w_i^{(n)}]=\lambda_nw_i^{(n)},
\end{equation} 
where $n=1,\ldots,N$ labels the mode. The axial
and transverse eigenfrequencies are given by the relations
$\omega_n^{\|~2}=\nu^2+\lambda_n$ and $\omega^{\perp
  2}_n=\nu_t^2-\lambda_n/2$. Since $\lambda_n>0$, 
the axial collective excitations are of higher frequencies
than the trap frequency $\nu$, while the transverse excitations are of lower
frequencies than $\nu_t$. In general, Eqs.~(\ref{Eq:ax}-\ref{Eq:t})
are invariant under reflection with respect to the center of the trap, which
coincides with the origin of the axes. Hence, the normal modes of the
chain are symmetric (even) or antisymmetric (odd) under reflection with
respect to the center, $w_i=\pm w_{N-i}$~\cite{Kielpinski}. 
It can be easily verified that the
center-of-mass motion is an even eigenmode, $w^{(1)}$, 
of the Eqs.~(\ref{Eq:ax}) and~(\ref{Eq:t}) at frequency $\nu$, $\nu_t$, 
respectively. Remarkably, also
the first odd eigenmodes can be exactly determined. They are $w_i^{(2)}\propto
x_i^{(0)}$ at the eigenfrequencies $\sqrt{3}\nu$ for the axial,
$\sqrt{\nu_t^2-\nu^2}$ for the transverse excitation, independent of the 
number of ions, as can be found by substitution of this ansatz in
Eq.~(\ref{I}-\ref{I:2}), 
combined with the equilibrium condition~\cite{Morigi}.

\begin{center}
\begin{figure}
\epsfxsize=0.25\textwidth
\epsffile{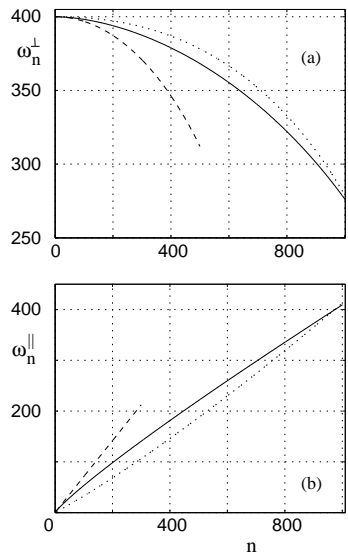}
\caption{(a) Transverse and (b) axial spectrum. The eigenfrequencies are in
  units of $\nu$. Here, $N=1000$ and 
$\nu_t=400\nu$. Solid line: numerical solution of
Eqs.~(\ref{Eq:ax}-\ref{I}). Dashed line:
 (a) $\omega_n^{\perp~{\rm Jac}}$, (b) $\omega_n^{\|~{\rm Jac}}$: These curves
 have been truncated, as they do not correctly 
reproduce the short wavelength eigenmodes. Dashed-dotted line: spectra
obtained from the density of states~(\ref{Density:Modes}). }
\label{Fig:1}
\end{figure}
\end{center}

\begin{center}
\begin{figure}
\epsfxsize=0.5\textwidth
\epsffile{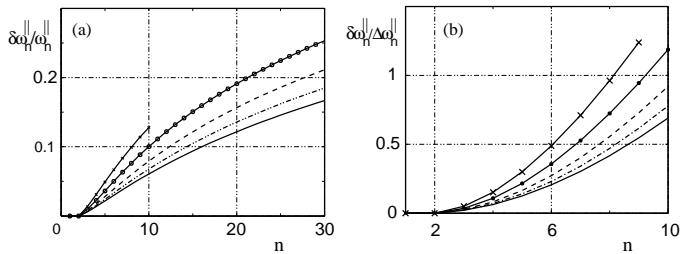}
\caption{(a) $\delta\omega_n^{\|}/\omega_n^{\|}$ 
and (b) $\delta\omega_n^{\|}/
  \Delta \omega_n^{\|}$
as a function of $n$, where 
$\delta\omega_n=\omega_n^{\|{\rm Jac}}-\omega_n^{\|}$ and
$\Delta\omega_n^{\|}=\omega_{n+1}^{\|}-\omega_n^{\|}$. 
The frequencies $\omega_n^{\|}$ are obtained by solving
numerically~(\ref{Eq:ax}-\ref{I}). 
From top to bottom: $N=10, 50, 200, 500, 1000$.} \label{Fig:0}
\end{figure}
\end{center}

In order to solve Eq.~(\ref{I:2}), we assume $N\gg 1$. In this limit,
we may consider the interparticle spacing 
$a_L(x_i)=x_{i+1}^{(0)}-x_i^{(0)}$ a smooth function of the position, 
which is inversely proportional to the density 
of ions per unit length $n_L(x_i)=1/a_L(x_i)$. The density $n_L(x)$ is
determined 
assuming a uniform distribution of charges inside an oblate ellipsoid
of axial length $2L$, and takes the form $n_L(x)=3N/4L\left(1-x^2/L^2\right)$,
which is defined for $|x|\le L$, while the length $L$ is found
by minimizing the energy of the crystal, and
fulfills the relation $L(N)^3=3\left(Q^2/m\nu^2\right)N\log N$~\cite{Dubin97}. 
These approximations provide a good estimate of the ions distribution in the
center of the chain for $N$ sufficiently large. In fact, $n_L(x)$
and $L(N)$ used here are the leading terms of an expansion in powers of
$1/\log N$~\cite{Dubin97}. 
For $N\gg 1$ one can approximate the chain by a continuous
distribution of charges for the evaluation of the
long-wavelength modes. In this limit, $x_i^{(0)}$ is a 
continuous variable in the interval $(-L,L)$, the displacement 
$w_i= w(x_i^{0})$ is a continuous function denoted by $w(x)$, and the sum
in~(\ref{I}) is approximated by an integral, where the density $n_L(x)$
appears as a weight function. In particular, the eigenmodes fulfill the
orthogonality relation $\int_{-L}^L{\rm d}x n_L(x) 
w^{(n)}(x)^{*}w^{(m)}(x)=\delta_{n,m}$. Using
an integration by parts the sum~(\ref{I}) can be approximated by
\begin{eqnarray}
I[x,w(x)]\approx -{\cal I}_0[(1-\xi^2)w^{\prime\prime}(x)-4\xi
w^{\prime}(x)] 
\label{Jacobi}
\end{eqnarray}
where $\xi=x/L$ and ${\cal I}_0=3NQ^2/(2mL^3)\approx
\nu^2/2$~\cite{Morigi}. The approximation
is valid at leading order in $\log N$ and sufficiently
far away from the edges of the chain. The eigenfunctions of Eq.~(\ref{Jacobi})
are the Jacobi polynomials $P^{1,1}_{\ell}(\xi)$~\cite{Abramowitz} 
with the eigenvalues $\lambda_n^{\rm Jac}=-\ell(\ell+3)\nu^2/2$ with
$\ell=0,1,\ldots$ and $n=\ell +1$. Substituting this result
into the eigenvalue equations~(\ref{I}-\ref{I:2}), one finds  
\begin{equation}
\label{omega:ax}
\omega_n^{\|~{\rm Jac}}=\nu\sqrt{n(n+1)/2}.
\end{equation}
Analogously, the eigenfrequencies of the transverse modes are
$\omega_n^{\perp~{\rm Jac}}=\sqrt{\nu_t^2-\nu^2 (n-1)(n+2)/4}$. 
It should be noted that the Jacobi polynomials at $\ell=0,1$ are exact 
solutions
of Eqs.~(\ref{Eq:ax}-\ref{Eq:t}). In general, they are the correct
eigenmodes of Eq.~(\ref{I:2}) in the asymptotic limit $N\to\infty$. For finite
$N$ a perturbative  
expansion in the parameter $1/\log N$ is required, resulting in
a very slow convergence. Nevertheless, the behaviour
$\lambda_{n}\propto\nu^2$ is exact in the continuum 
limit~\cite{Dubin97,Morigi}.  
Figure~\ref{Fig:1} presents the comparison between the spectrum of 
eigenfrequencies, calculated numerically for 1000
ions, and the solutions $\omega_n^{\|~{\rm Jac}}$, $\omega_n^{\perp~{\rm
    Jac}}$. Figure~\ref{Fig:0} shows the relative
deviation of the frequencies~(\ref{omega:ax}) from the numerical
results for chains of different number of ions. From Fig.~\ref{Fig:0}(a)
one sees that Eq.~(\ref{omega:ax}) gives already a good estimate of the
eigenfrequencies for a chain of 10 ions. Its prediction is
valid for the axial low- and the transverse high-frequencies, and it improves
slowly as N increases, as expected from the slow convergence of the
$1/\log N$ expansion. 
\vskip -1.7cm

\begin{center}
\begin{figure}
\epsfxsize=0.25\textwidth
\epsffile{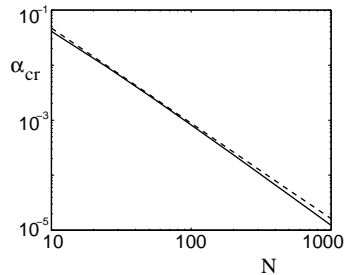}
\caption{$\alpha_{\rm cr}(N)$ as a function of the number of ions. The
  solid curve gives the result according to~(\ref{alpha}). The
  dashed line is the curve $2.53~N^{-1.73}$, that fits the values obtained by
molecular dynamics simulations~\protect\cite{Schiffer93,LosAlamos00}.}  
\label{Fig:Stable}
\end{figure}
\end{center}

The short wavelength eigenmodes are
characterized by relatively large displacements of the ions at the chain
center, while the ions at the edges nearly do not move. In
fact, the interparticle distance is minimal around the center, 
and for $N\gg 1$ it is of order $1/N$, while it is significantly
larger at the edges. Hence, a wave cannot propagate to the edges, where the
interparticle spacing is larger than the wavelength. Moreover, in the center
and for the short-wavelength excitations we expect that 
the relevant contributions to the force
on an ion originate from the neighbouring charges, as 
nearby groups of ions move in opposite directions, resulting in
forces that mutually cancel. By this hypothesis
we keep in~(\ref{I}) only the nearest--neighbour interaction
and apply the formalism developed by Dyson~\cite{Dyson53} to 
derive the density of states as a function of the squared frequencies
$\lambda$. This is given by
\begin{eqnarray}
D(\lambda)=8/(\pi N\lambda)\int_{0}^{f(\lambda)}{\rm
  d}x~n_L(x)\left[\sqrt{-1+4\Lambda(x)/\lambda}\right]^{-1}
\label{Density:Modes}
\end{eqnarray}
where $\Lambda(x)=2Q^2n_L(x)^3/m$ is the nearest-neighbour coupling constant
for slowly-varying interparticle distance, and
$f(\lambda)=L\sqrt{1-(\lambda/4\Lambda(0))^{1/3}}$~\cite{Morigi}. Here,
$\Lambda(0)\approx 9 \nu^2N^2/32\log N$ at leading order in
$\log N$. The spectrum obtained from Eq.~(\ref{Density:Modes}) is shown in
Fig.~\ref{Fig:1}. The deviations from the numerical result are due to the
assumption of nearest-neighbour coupling, and are small in the
short-wavelength part of the spectrum, showing that Eq.~(\ref{Density:Modes})
provides a good approximation in this regime. In particular,
it provides a good estimate of the eigenvalue $\lambda_N$, which
determines the maximal axial- and minimal transverse-frequencies through the
relations $\omega_{\rm max}^{\|}=\sqrt{\nu^2+\lambda_N}$ and 
$\omega_{\rm min}^{\perp}=\sqrt{\nu_t^2-\lambda_N/2}$. This
eigenvalue is found from $f(\lambda)=0$, as for
$\lambda>\lambda_N$ the function $D(\lambda)$ vanishes, and
$\lambda_N=9\nu^2N^2/(8\log N)$ at leading order in $\log N$.
The minimal transverse frequency is
\begin{equation}
\label{w:min:t}
\omega_{\rm min}^{\perp}\approx\sqrt{\nu_t^2-
\frac{9}{16}\frac{\nu^2N^2}{\log N}}.
\end{equation}
It can vanish for certain values of $\nu,\nu_t$ and $N$. We identify  
the critical value $\nu_{t,{\rm cr}}=\sqrt{\lambda_N/2}$, 
such that for $\nu_t<\nu_{t,{\rm cr}}$ the linear chain is unstable.
The critical aspect ratio
between the trap frequencies $\alpha_{\rm cr}=\nu^2/\nu_{t,{\rm cr}}^2$ takes
the value  
\begin{equation}
\label{alpha}
\alpha_{\rm cr}=\frac{16}{9}\frac{\log N}{N^2}
\end{equation}
which fixes the condition on $\nu_t$ for the chain stability 
according to the inequality
$\nu_t^2>\alpha_{\rm cr}^{-1}\nu^2$. The result~(\ref{alpha}) 
is in agreement with the analytical
estimate in~\cite{Dubin93}, which was obtained under different requirements. 
Figure~\ref{Fig:Stable} shows that it also approximately reproduces the curve
obtained by fitting points calculated 
by molecular dynamics simulation~\cite{Schiffer93}, which has been
experimentally verified for chains of tens of ions~\cite{LosAlamos00}.
It was suggested that this instability, resulting 
in a structural transition from a linear to a zig-zag equilibrium
configuration~\cite{MPQ,Schiffer93}, can be treated as a phase
transition~\cite{Schiffer93,LosAlamos00}. 
\vskip -0.8cm

Using these results we now discuss the thermodynamics of a linear
crystal of $N$ ions, whose dynamics are
described by $3N$ harmonic oscillators of frequencies $\omega_n^{\|}$,
$\omega_n^{\perp}$, in the regime where the axial frequencies are 
much smaller than the transverse excitations, i.e.\ $\omega_{\rm max}^{\|}\ll
\omega_{\rm min}^{\perp}$. The eigenmodes are quantized using standard
procedures~\cite{James}.  The chain is assumed to be in thermal
equilibrium with a bath at temperature $T$, which is sufficiently low so that
only the axial modes are excited ($k_BT\ll \hbar\omega_{\rm max}^{\perp}$). 
Hence, the dynamics of the system is one-dimensional. 
The thermodynamic limit for this kind of system can be 
defined by assuming constant interparticle spacing (thus constant 
linear density) at the center of the chain as $N$ increases, namely
requiring $a_L(0)$ to be constant, in analogy with the definition for 
cold gases in traps~\cite{Stringari}. Since
$a_L(0)\propto\left(\sqrt{\log N}/\nu N\right)^{2/3}$, this 
requirement corresponds to a vanishing axial frequency as $N\to\infty$,
according to $\nu\sim\sqrt{\log N}/N$. Using this definition, $\omega_{\rm
  mac}^{\|}$ and $\omega_{\rm min}^{\perp}$ take finite values in the
thermodynamic limit, since they depend  on $N$ and $\nu$
only through the combination $\nu N/\sqrt{\log N}$.
We identify the thermodynamic variables with $T$, $N$, and $\nu$, whose variation results in a variation in the  
length of the chain~\cite{DubinRMP}, and take constant $\nu_t$. 
First we study
the heat capacity $C_a=\partial U_{\rm th}/\partial T|_{\nu,N}$, where the
thermal energy $U_{\rm th}$ is the average energy of the excited state
relative to the ground state energy. At high temperatures $C_a$
exhibits the Dulong--Petit law $C_a=Nk_B$, while at low
temperatures it takes the form
 \begin{equation} 
C_a\sim \frac{\sqrt{2}k_B^2}{\hbar\nu}
\frac{\partial}{\partial T}T^2
\int_{x_0}^{\infty} {\rm d}x \frac{1}{{\rm e}^{x}-1}\frac{x^2}
{\sqrt{x^2+x_0^2/8}}
\label{cV:1}
\end{equation}
where $x_0=\beta\hbar\nu$ and we have used the density of states derived
from~(\ref{omega:ax}). For $k_BT\gg\hbar\nu$ 
we can set $x_0\sim 0$ and obtain 
$C_a\propto T$. Hence, in this temperature regime the heat capacity is linear
in $T$, like for a one-dimensional Debye crystal~\cite{Ashcroft}. 
Remarkably, for these temperatures $C_a$ is inversely proportional to $\nu$. 
Hence, the specific heat per
particle $c_a=C_a/N$ behaves like $c_a\sim 1/N\nu$ 
and vanishes as $c_a\sim 1/\sqrt{\log N}$
in the thermodynamic limit defined above. Thus,
at low temperatures it depends on the number of ions, thereby manifesting a
deviation from extensivity of the system's behaviour. \\
\indent The coefficient of thermal expansion
$\alpha_T$ also exhibits a remarkable behaviour as a function of
the temperature. 
It is related to $C_a$ and to the isothermal compressibility
$\kappa_T$ according to the relation  
$\alpha_T=3\kappa_TC_a/2L$~\cite{Ashcroft}. In the regime of thermal
stability, the isothermal compressibility does not vanish and 
is practically independent of the temperature, the pressure being dominated by
the zero-temperature component. In particular, it scales as $\kappa_T\sim
1/\log N$. Hence, the temperature dependence of $\alpha_T$ is solely
determined by the heat capacity $C_a$. At low temperatures $C_a\sim 1/\nu$,
and as the thermodynamic limit is approached $\alpha_T\sim (\log
N)^{-3/2}$. At higher  temperatures, when the heat capacity manifests the
Dulong--Petit behaviour $C_a=Nk_B$, the coefficient of thermal expansion
vanishes like  $\alpha_T\sim 1/\log N$. For large but finite number of ions
$\alpha_T$ does not vanish, in contrast to the behaviour found in a
uniform harmonic solid~\cite{Ashcroft}. The coefficient of thermal expansion 
$\alpha_T$ is plotted in Fig.~4 as a
function of $T$ for a chain of 1000 ions.
\begin{center}
\begin{figure}
\epsfxsize=0.25\textwidth
\epsffile{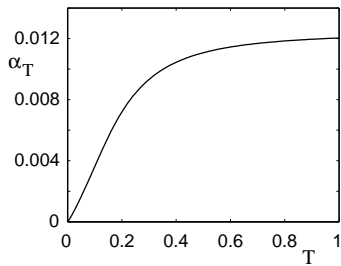}
\caption{Coefficient of
  thermal expansion $\alpha_T$, in arbitrary units, as a function of 
$T$, in units of $\hbar\omega_N^{\|}/k_B$. Here, $N=1000$ and 
$\omega_{\rm max}^{\|}\ll\omega_{\rm min}^{\perp}$. This would correspond to
an experiment with $\nu=2\pi\times 1$ kHz and $\nu_t=2\pi\times 3$ MHz. }
\label{Fig:4}
\end{figure}
\end{center}
\vskip -0.8cm
\indent 
Finally, we remark on the regime
of thermal stability, which holds
when the thermal energy is considerably smaller than the
equilibrium energy, or, equivalently, the displacements are much
smaller than the respective interparticle distances. The
condition for thermal stability is $Q^2\log N/a_L(0)\gg k_BT$. 
Outside this regime,
thermal excitations may cause structural
transitions, accompanied by a critical behaviour of the thermodynamic
functions. 
 
In conclusion, we studied the low temperature dynamics 
of a strongly-correlated and one-dimensional crystal inside a harmonic
potential. We investigated how the dynamics of the system is affected by the
dimensionality and the long-range correlations and how this is manifest in the
thermodynamics functions. In future works we will explore the behaviour of
these functions at the instability points, in order to characterize the
crystal structural transitions and their relations to the standard theory of
phase transitions~\cite{Lubensky}.   

The excitations spectra we presented are of interest for spectroscopy and
quantum information with ion traps, where quantum logic is 
implemented with the lowest collective modes~\cite{Cirac95,Steane,QLogic:2}.
This work may be relevant to studies of Coulomb
systems like ion crystals in storage rings~\cite{Habs} and cold neutral plasmas~\cite{GallagherReview}, which at sufficiently low
temperatures are expected to crystallize~\cite{Pattard}, and contributes to
the on-going research on low-dimensional cold gases~\cite{BEC1d}. 

It is our great pleasure to thank Andreas Buchleitner for hospitality at
MPIPKS in Dresden, where most part of the research was done, and J.\ Eschner
for careful reading of the manuscript.
Support by BSF, the Minerva Center of Nonlinear
Physics of Complex Systems, the Center for Theoretical Physics 
at the Technion, QUEST, and QGATES, is acknowledged.

\end{document}